\documentclass[]{spie}

\usepackage{amsmath}	
\usepackage{siunitx}
\usepackage{txfonts}
\usepackage{graphicx}
\usepackage[utf8]{inputenc}
\usepackage{graphicx}	
\usepackage{multirow}
\usepackage{amsfonts}
\usepackage{adjustbox}
\usepackage{placeins}

\usepackage{booktabs}
\usepackage{hyperref}       
\usepackage{bm}

\def\CN2{\mbox{$C_N^2 $}}

\begin{document} 

   \title{Establishing baseline model performances for optical turbulence forecasting}

    \author[a,b]{M. de Sepibus}
    \author[a,b]{E. Masciadri}
    \author[a,b]{C. Weinberger}
    \author[c]{A. Otarola}
    \author[c]{S. Mieske}
    \author[d]{C. Veillet}
    \author[d]{S. Ragland}
    
    \affil[a]{INAF Osservatorio Astrofisico di Arcetri, Largo Enrico Fermi 5, I-501 25 Florence, Italy}
    \affil[b]{ADONI: ADaptive Optics National laboratory of INAF}
    \affil[c]{European Southern Observatory, Alonso de Córdova, 3107 Vitacura, Stgo 763-0355, Chile}
    \affil[d]{Large Binocular Telescope, 933 N. Cherry Drive, Tucson, AZ, USA, 85721}
\authorinfo{marlene.desepibus@inaf.it, elena.masciadri@inaf.it}
\maketitle
  
\begin{abstract}

Accurate forecasting of optical turbulence (OT) is useful to optimise telescope scheduling and maximising observing efficiency, as the performance of adaptive optics systems strongly depends on turbulence conditions. While various numerical weather prediction and machine learning models are developed to predict these complex turbulence properties, properly evaluating their added value requires well-defined baseline reference models. This study establishes baseline performance thresholds for key astroclimatic parameters using two distinct models: one forecasting the parameter average over a specific timescale, and the other forecasting the temporal evolution over the same specific timescale. We apply these frameworks to extensive, multi-year datasets from the sites of the Very Large Telescope (VLT) and the Large Binocular Telescope (LBT) and discuss the possibility of extrapolating reference values of forecast performances for other sites based on their optical turbulence statistical characteristics. Our results show that baseline predictive accuracy varies according to the observing site, the specific parameter, the chosen timescale, and the forecasting approach. Therefore, an OT forecasting model's accuracy cannot be quantified in absolute terms, but relatively to the baseline.

\end{abstract}

\section{Introduction}
\label{intro}
This contribution presents an overview and selected results of the full study detailed in de Sepibus et al. \cite{desepibus2026}.
\\

Ground-based astronomical observations are fundamentally limited by atmospheric optical turbulence (OT), which degrades telescope image quality. While adaptive optics (AO) systems mitigate these effects, their performance remains highly dependent on OT conditions. Consequently, precise OT forecasting is essential for queue scheduling and proactive observation planning to maximise operational efficiency.

To date, various methodologies have been developed for OT forecasting, including numerical weather prediction models where the OT is parameterized ~\cite{masciadri1999,cherubini2008,masciadri2017,basu2020}, empirical methods~\cite{trinquet2007,giordano2013,osborn2018}, machine learning algorithms~\cite{milli2019,bolbasova2021,cherubini2022,masciadri2023,li2023}, and hybrid approaches that join the numerical framework with an autoregressive algorithm~\cite{masciadri2020,masciadri2023}.   Evaluating the added value these diverse models offer to telescopes requires matching them against simple, computationally inexpensive, baseline models.  If the accuracy of a forecasting model cannot surpass that of the baseline, it cannot be considered effective and useful. 

We developed two distinct baseline models for short-term predictions (on the order of 1–2 hours): Precasting (PC), which forecasts the average value of a parameter over a specific time interval, and Prevision by Persistence (PP), which forecast the temporal evolution over the same specific time interval.  We used multi-year statistical datasets from two astronomical sites: Cerro Paranal (Chile), site of the Very Large Telescope (VLT), and Mt. Graham (USA), site of the Large Binocular Telescope (LBT) and we quantified the performance of the PC and PP models for key astroclimatic parameters: seeing ($\varepsilon$), wavefront coherence time ($\tau_0$), isoplanatic angle ($\theta_0$), and the ground layer fraction (GLF). We refer the readers to the more extended version of this study \cite{desepibus2026} for the analysis performed for the atmospheric parameters.

The motivations driving this study are threefold: 
\begin{enumerate}
    \item To establish robust baseline reference thresholds using extensive multi-year datasets above two top-class observatories where operational forecast systems, more precisely the ALTA Center\footnote{\href{http://alta.arcetri.inaf.it}{http://alta.arcetri.inaf.it}} at LBT and FATE at VLT~\cite{masciadri2023}, are active and leaded by our team; 
    \item To identify which are the parameters from which these baseline performances depend. Possible parameters that might play a role include the geographic site, the astroclimatic parameter, and the defined forecast timescale;
    \item To determine whether baseline performance trends can be extrapolated to predict thresholds at alternative sites. To test the final hypothesis, we enriched our dataset by using observations of the seeing from a few further sites: Roque de Los Muchachos Observatory (Canary Islands) using the DIMM of the TNG~\cite{molinari2012,deguturbai2013}, La Silla Observatory (Chile) using a DIMM, Matera Space Centre Laboratory (Italy) using a Cyclope instrument~\cite{dicecco2023}, Cagliari (Italy), and Kryoneri (Greece) using a Cyclope. The latter three sites are not astronomical locations and they have been expressly selected as we expect there bad seeing, more precisely a worse seeing with respect to astronomical sites.  
\end{enumerate}
   

\section{Models and forecast timescale}
\label{def}

\subsection*{Baseline Forecasting Models}
\label{baselines}
\begin{itemize}
\item {\bf Prediction by Persistence (PP):} This baseline model forecasts the temporal evolution of a parameter by assuming the variable remains constant over the targeted forecast window. We study whether we can obtain a more accurate value by taking the average of observations over a prior look-back interval ($\Delta t$) instead of the value at time ($t_0$). The optimal value for $\Delta t$ is determined by minimising the root-mean-square error (RMSE) against real measurements. To preserve the trend at low frequencies and to eliminate high-frequency temporal fluctuations that are irrelevant for our application the verification observations are processed, after the calculation of the $\Delta t$ with a 1-hour centred moving average.
\item {\bf Precasting (PC):} This baseline model is used to forecast the mean value of a parameter over an interval of time. It calculates the average of the observations recorded during a prior look-back interval ($\Delta t^*$), and it assumes it is the expected average value for forecast time scale of interest. The optimal duration for $\Delta t^*$ is determined using the same RMSE minimisation approach as the PP model. Unlike PP, verification observations for PC do not undergo moving average processing because the model evaluates a scalar average rather than a continuous temporal evolution.
\end{itemize}

\subsection*{Forecast Timescale (FTS)}
\label{fts_def}
\begin{figure}[h]
\centering
\includegraphics[height=5cm]{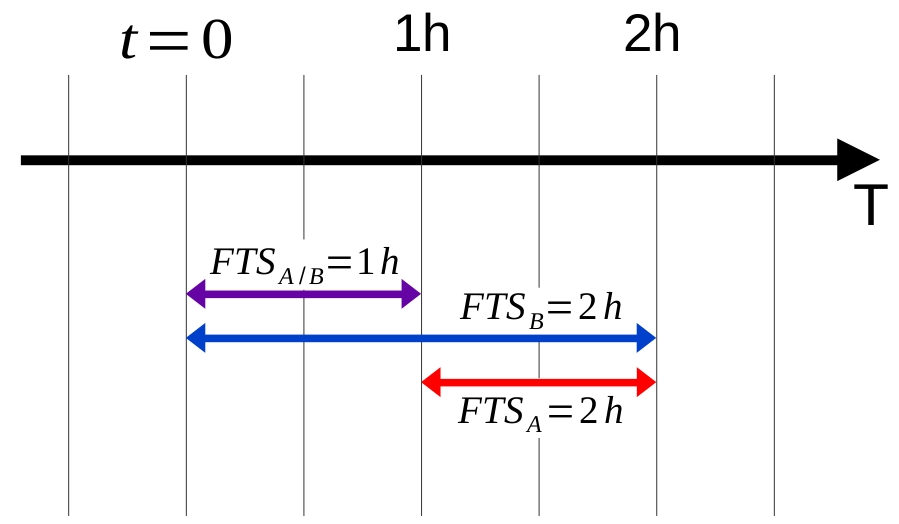}
\caption{Schematised representation for cases (A) and (B) of FTS related to 1h and 2h. The arrows indicate the start/end of each interval} 
\label{FTS_defintion}
\end{figure}

For details, refer to \cite{desepibus2026}.
{ Breifly, the FTS defines the interval of time between the time at which the initialisation forecast is calculated and the time at which the forecast refers to.}  Because both baseline models rely strictly on real-time observations, forecasts are generated at the present time ($t = 0$). {We provide here two distinct FTS definitions} (schematized in Figure~\ref{FTS_defintion}):
\begin{itemize}
\item {\bf Case (A):} { The FTS of $X$ hours (where X can be 1h, 2h etc.), is the  interval $[t_0 + X - 1, t_0 + X]$}
\item {\bf Case (B):} { The FTS of $X$ hours (where X can be 1h, 2h etc.), is interval $[t_0, t_0 + X]$.}
\end{itemize}

By definition, Case (A) and Case (B) are identical for a 1-hour forecast { timescale}. For longer horizons, Case (B) naturally shows a smaller average error because it includes the hours closest to the present, which are easier to predict.

\section{Observations}
\label{obs}
For this study, we used observations of optical turbulence parameters relevant for ground-based astronomy at seven sites: (1) Cerro Paranal (Chile), the site of the Very Large Telescope (VLT), (2) Mt. Graham (US), the site of the Large Binocular Telescope (LBT), (3) Roque de los Muchachos Observatory (La Palma, Canary Islands) the site of the Telescopio Nazionale Galileo (TNG), (4) La Silla (Chile) an ESO site, with the New Technology Telescope (NTT), (5) the Matera Space Center Laboratory from the Italian Space Agency (ASI) close to Matera (Italy), (6) Cagliari (Italy) and (7) Kryoneri (Greece). The main analysis is focused on the first two sites. The last five sites are used for a specific analysis presented in Section \ref{disc}.

For the VLT, we had access to measurements from a MASS-DIMM (Multi Aperture Scintillation Sensor, Differential Image Motion Monitor), ranging 8 years from 2017 to 2024. For the LBT and TNG, we had access to measurements from a DIMM, ranging 8 years from 2017 to 2024. For the NTT, we had access to measurements from a DIMM, ranging 7 years from 2017 to 2023. For Matera, we had access to Cyclops measurements for the year 2022. For Cagliari and Kryoneri, we had access to data from a C-DIMM for the years 2024 and 2025. 

We highlight that the cases of Matera, Cagliari and Kryoneri are statistically less representative, as the sample covers 1 or 2 years instead of 8 years. { This will be discussed in Section \ref{disc}.}


\section{Analysis}
\label{ana}

Readers are referred to de Sepibus et al \cite{desepibus2026} for the explicit mathematical formulations and the full analysis of the preliminary data processing elements.

\begin{figure}[h]
\centering
\includegraphics[height=7cm]{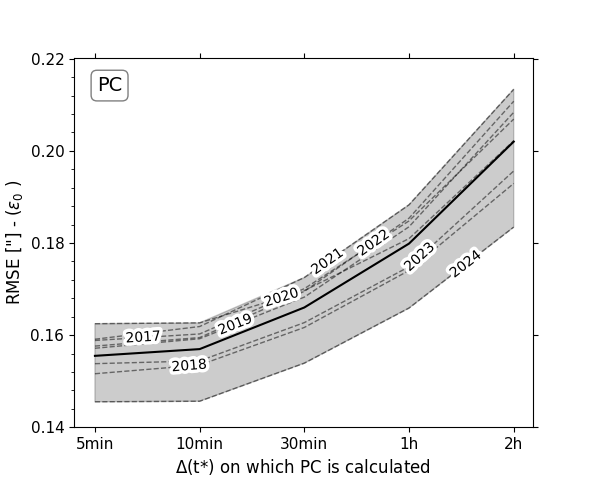}
\caption{RMSE between observations and precasting (PC) above Paranal for a FTS = 1h as a function of the interval of time $\Delta(t^*)$ over which the PC is calculated. The bold black line is the RMSE obtained with the whole sample of 8 years. The dashed lines report RMSE obtained in each calendar year. The grey area represents the uncertainty.} 
\label{criterium}
\end{figure}

To quantify forecasting performance, the root-mean-square error (RMSE) is used as the figure of merit for both Precasting (PC) and Prevision by Persistence (PP). A dispersion band spanning annual results tracks the variability around the multi-year mean RMSE (see Figure~\ref{criterium}). Analysis of the raw sampling frequencies shows that the statistical properties of the turbulence are preserved up to a resampling interval of 1.5--2~minutes; hence, the datasets from the different observatories are homogeneous and comparable { (see Fig.3 de Sepibus et al.\cite{desepibus2026}).}

We studied forecast performances using our baseline models { for} the main available astroclimatic parameters: total seeing ($\varepsilon$), truncated seeing ($\varepsilon < 1.5$ arcsec), wavefront coherence time ($\tau_0$), isoplanatic angle ($\theta_0$), and ground layer fraction (GLF). The $\varepsilon < 1.5$ arcsec threshold provides an operational range where adaptive optics (AO) performance can be guaranteed, as AO systems cannot typically be used under high-turbulence conditions. Due to the lack of vertical optical turbulence profilers at Mt. Graham, $\tau_0$, $\theta_0$ and GLF cannot be calculated for the LBT site.

\begin{figure}[h]
  \centering
  \includegraphics[width=0.9\hsize]{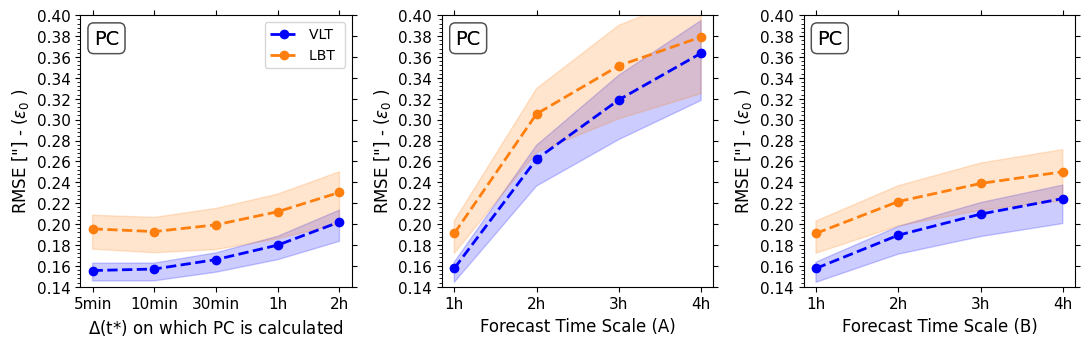}
  \caption{Seeing: RMSE between observations and precasting (PC) above Paranal(blue line) and Mt.Graham (orange line) using the dataset belonging to the 8 year range [2017,2024]. Left: RMSE as a function of the interval of time  $\Delta(t^*)$  on which the PC is calculated for FTS = 1-hour. Centre: RMSE as a function of the FTS~-~CASE (A). Right: RMSE as a function of the forecast time scales for the CASE (B). In the last two columns, for each parameter, the optimal  $\Delta(t^*)$ (10min) is retrieved from the first column.}
  \label{seeing_PC}
\end{figure}

\begin{figure}
  \centering
  \includegraphics[width=0.9\hsize]{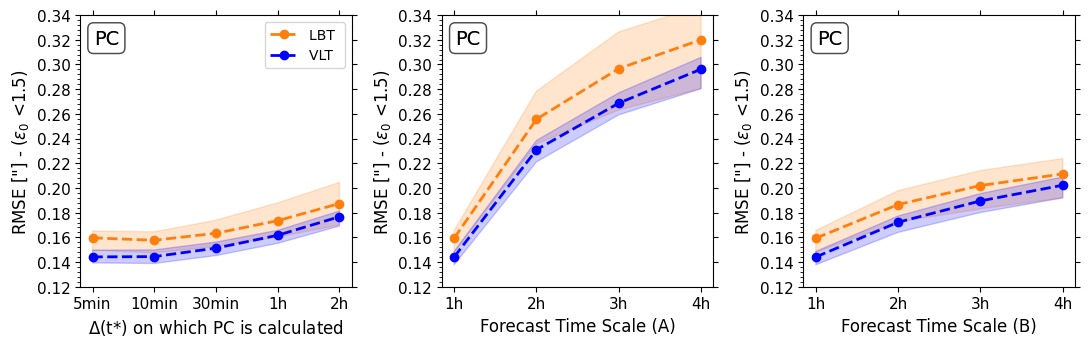}
  \caption{Same as Fig.\protect\ref{seeing_PC} but for the $\varepsilon$ < 1.5 arcsec.  That means that when we calculate the RMSE, we consider only the couples in which the element related to observations is below 1.5 arcsec. The optimal  $\Delta(t^*)$  is 10 minutes  }
  \label{seeing_15_PC}
\end{figure}

\begin{figure}
  \centering
  \includegraphics[width=0.9\hsize]{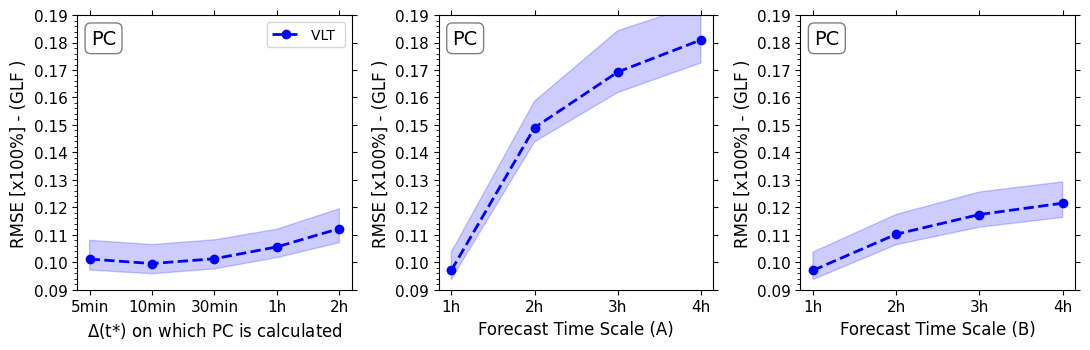}
  \caption{Same as Fig.\protect\ref{seeing_PC} but for the Ground Layer Fraction. The optimal  $\Delta(t^*)$  is 10 minutes.}
  \label{GLF_PC}
\end{figure}

\begin{figure}
  \centering
  \includegraphics[width=0.9\hsize]{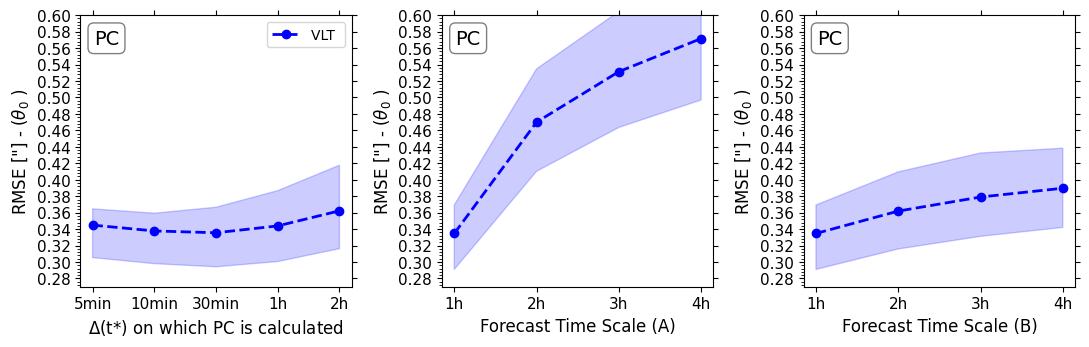}
  \caption{Same as Fig.\protect\ref{seeing_PC} but for the  isoplanatic angle. The optimal  $\Delta(t^*)$  is 30 minutes}
  \label{Theta_PC}
\end{figure}

\begin{figure}
  \centering
  \includegraphics[width=0.9\hsize]{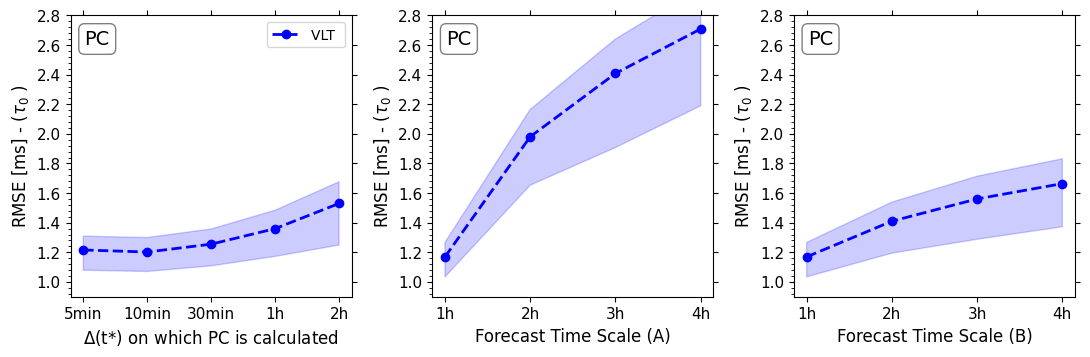}
  \caption{Same as Fig.\protect\ref{seeing_PC} but for the  wavefront coherence time. The optimal  $\Delta(t^*)$  is 10 minutes.}
  \label{Tau_PC} 
\end{figure}

\begin{figure}
  \centering
  \includegraphics[width=0.9\hsize]{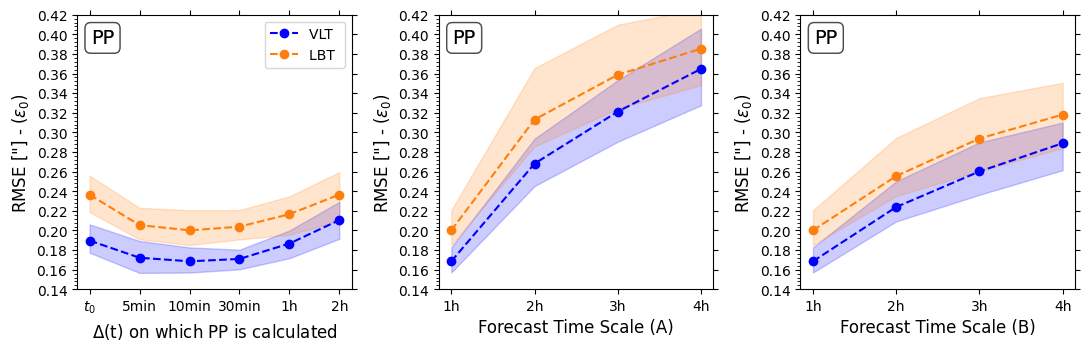}
  \caption{Seeing: RMSE between observations and prediction by persistence (PP) above Paranal(blue line) and Mt.Graham (orange line) using the dataset belonging to the range of 8 years [2017,2024]. Left: RMSE as a function of the interval of time $\Delta$t on which the PP is calculated for FTS = 1-hour. Centre: RMSE as a function of the forecast time scales~-~CASE (A). Right: RMSE as a function of the forecast time scales for CASE (B). In the last two columns we use, for each parameter, the optimal $\Delta$t (10min) retrieved from the first column. See text in Section \protect\ref{ana}.}
  \label{seeing_PP}
\end{figure}

\begin{figure}
  \centering
  \includegraphics[width=0.9\hsize]{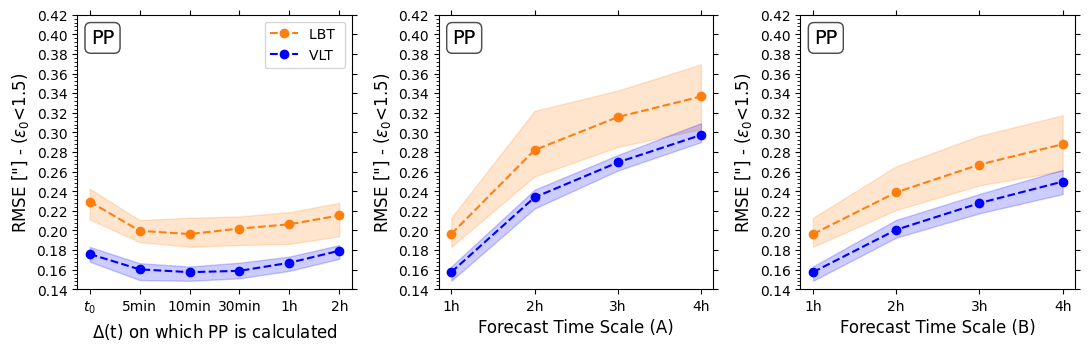}
  \caption{Same as Fig.\protect\ref{seeing_PP} but for the $\varepsilon$ < 1.5 arcsec. The optimal  $\Delta(t)$ is 10~min.}
  \label{seeing_15_PP}
\end{figure}

\begin{figure}
  \centering
  \includegraphics[width=0.9\hsize]{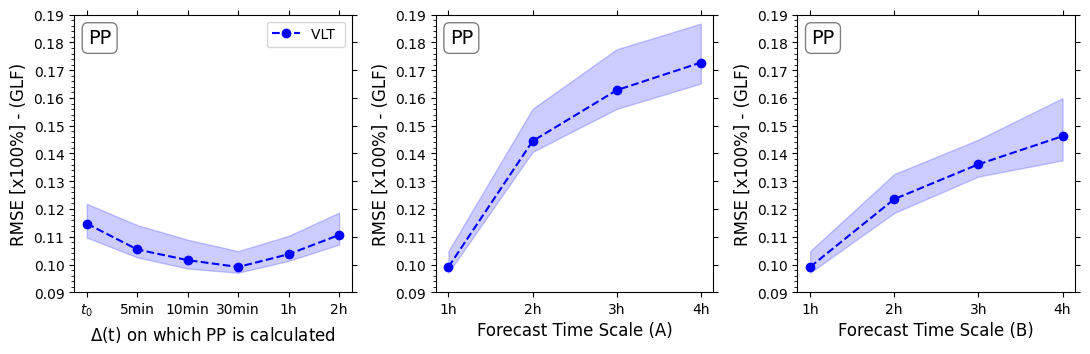}
  \caption{Same as Fig.\protect\ref{seeing_PP} but for the Ground Layer Fraction. The optimal  $\Delta(t)$ is 30~min.}
  \label{GLF_PP}
\end{figure}

\begin{figure}
  \centering
  \includegraphics[width=0.9\hsize]{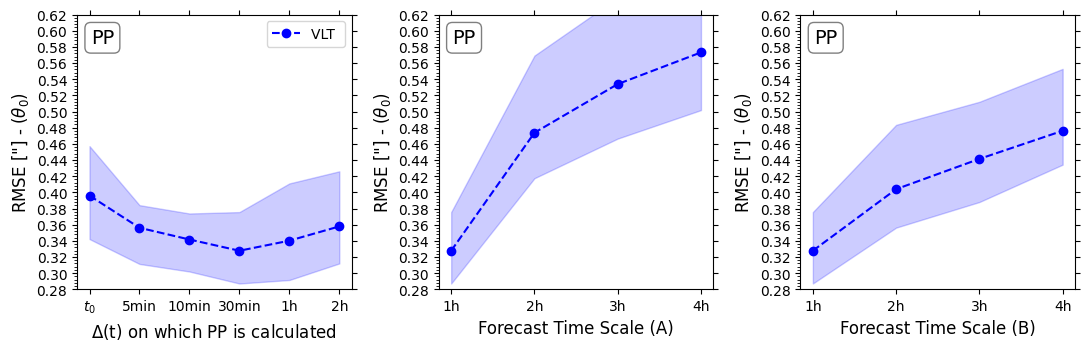}
  \caption{Same as Fig.\protect\ref{seeing_PP} but for the  isoplanatic angle. The optimal  $\Delta(t)$ is 30~min}.
  \label{Theta_PP}
\end{figure}

\begin{figure}[h]
  \centering
  \includegraphics[width=0.9\hsize]{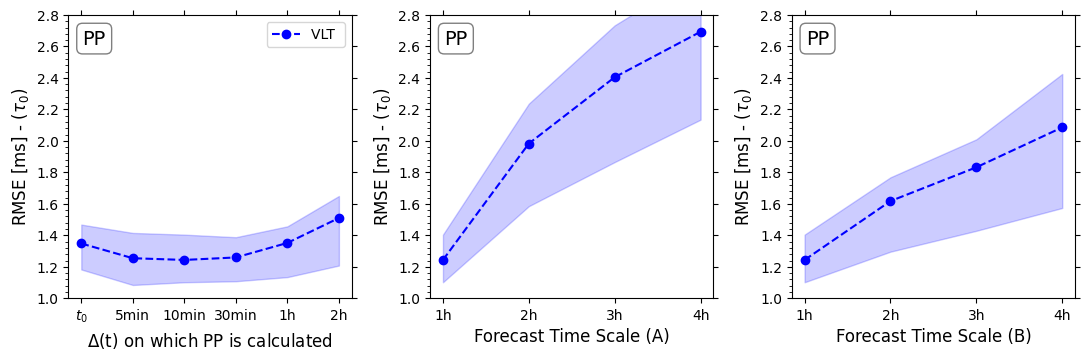}
  \caption{Same as Fig.\protect\ref{seeing_PP} but for the  wavefront coherence time. The optimal  $\Delta(t)$ is 10~min.}
  \label{Tau_PP} 
\end{figure}

For each parameter, the optimal look-back windows ($\Delta t^*$ for PC and $\Delta t$ for PP) were determined by minimising the RMSE for an FTS of 1 hour. As can be seen in Figures \ref{seeing_PC} to \ref{Tau_PP} in the left column, an optimal window of 10 minutes was identified for the majority of the astroclimatic parameters for both PP and PC. Exceptions were found for $\theta_0$ (PC and PP) and GLF (PP), which show an optimum at 30 minutes, though the impact on the RMSE is negligible. We refer the reader to { de Sepibus et al.} \cite{desepibus2026} for { the explanation and annexed} figures detailing these differences.

As seen in Figures \ref{seeing_PC} to \ref{Tau_PP}, the predictive accuracy for both PC and PP decreases with the increase of the FTS from 1 to 4 hours. For an FTS exceeding 1 hour, case (A) provides worse performance than case (B) by construction. This deterioration is sharper in case (A) because case (B) is influenced by the better predictive values close to $t_0$, while case (A) isolates values farther from $t_0$.

Comparing the two sites, the baseline performances for $\varepsilon$ and $\varepsilon < 1.5$ arcsec are consistently better at Paranal than at Mt. Graham across all evaluated forecast timescales. 

The values established for the operational 1-hour and 2-hour windows serve as the standard reference thresholds for the VLT and LBT sites. Any forecasting model accuracy must overcome these specific baselines using the identical configuration (FTS definition, case A or B, and data processing) to show it gives usefull information. The reader is referred to Table~1 in { de Sepibus et al.}\cite{desepibus2026} for the complete table summarising the numerical results for { all the} astroclimatic parameters { analyzed}.


\section{Discussion}
\label{disc}

\begin{figure}[h]
\centering
\includegraphics[width=0.9\hsize]{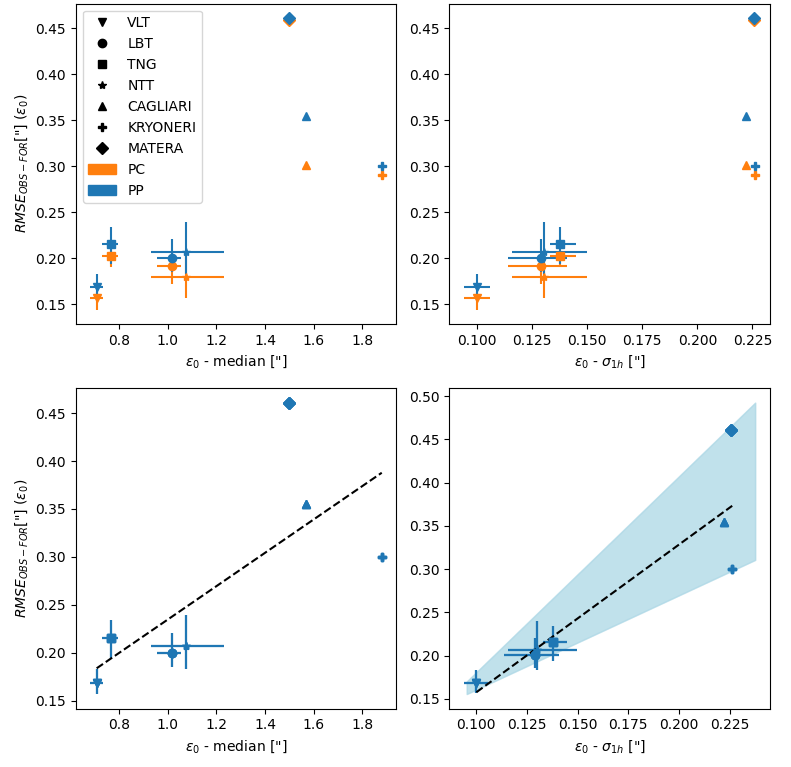}
\caption{Top: scattering plot between the RMSE$_{OBS-FOR}$ of the seeing with the median (left) and median hourly std (right) of the historical data for PP and PC. FTS = 1h, $\Delta t$ = 10min. Bottom: for the PC case, the black dashed line indicates the regression line between the RMSE$_{OBS-FOR}$ and the median (left) and the median  $\sigma_{1h}$ (right). The case for PP is equivalent. }
\label{RMSE_vs_mean_sigma}
\end{figure}

In this section, we investigate the possibility of finding a correlation between the statistical properties of the seeing and the $RMSE_{\text{OBS-FOR}}$ obtained for the PC and PP methods in Section~\ref{ana}. Our objective is to determine if a general law exists that allows us to retrieve reference threshold values for other locations based on easily identifiable OT characteristics. For this analysis we expand our sample to a total of seven sites: four astronomical sites (VLT, LBT, TNG (La Palma), NTT (La Silla, Chille)), and three non-astronomical sites characterized by less favourable conditions for the optical turbulence (Matera (Italy), Cagliari (Italy), and Kryoneri (Greece))  (see Section \ref{obs} for more infos on these sites).

 Figure \ref{RMSE_vs_mean_sigma} shows the $RMSE_{\text{OBS-FOR}}$ for both the PC and PP models versus two statistical operators: the median seeing and the standard deviation calculated on a 1-hour timescale ($\sigma_{1\text{h}}$). Results displayed in Figure~\ref{RMSE_vs_mean_sigma}, indicate a much better correlation between $RMSE_{\text{OBS-FOR}}$ and  $\sigma_{1\text{h}}$ than the median. More precisely, the $\sigma_{1\text{h}}$ provides a higher Pearson correlation coefficient and a lower RMSE relative to the regression line with respect to the $RMSE_{\text{OBS-FOR}}$ than the median seeing.  This indicates that the baseline predictive accuracy is not primarily governed by the typical seeing value of a site, but rather by the intrinsic variability of the turbulence on the same timescale over which the forecast error is computed.

We also observe two distinct behavioural regimes; for {$\sigma_{1\text{h}} < 0.15$ arcsec} the correlation is significantly tighter, indicating a good relationship between short-term stability and baseline error thresholds. While for $\sigma_{1\text{h}} > 0.15$ arcsec, the data-points are much more dispersed, as highlighted by the blue shadow in Figure~\ref{RMSE_vs_mean_sigma} (bottom right). 

We note important aspects regarding the data from Matera, Cagliari, and Kryoneri: their error bars appear artificially smaller because they are calculated from smaller samples (one or two years instead of eight years). With a full eight-year sample, we expect these less favourable sites would exhibit larger error bars { because the sample is larger and because they are characterized by a worse seeing than astronomical sites.} Furthermore, the pre-processing and cleaning of the measurements from these locations had a big impact on both $\sigma_{1\text{h}}$ and $RMSE_{\text{OBS-FOR}}$. 

Ultimately, while we cannot yet retrieve an explicit analytical expression for the regression law at present, we can state that a larger $\sigma_{1\text{h}}$ is systematically associated with a larger $RMSE_{\text{OBS-FOR}}$. Consequently, the baseline reference thresholds are much smaller at stable sites. This implies that providing a useful forecast (i.e., overcoming the baseline threshold) is significantly more difficult for highly stable astronomical sites. Conversely, these results show that optical turbulence forecasting is easier for applications in the Free-Space Optical Communications (FSOC), where systems are frequently operating at low altitudes or near urban areas with sub-optimal seeing conditions.

\begin{table}[h]
\caption{Values of different figures of merit quantifying the correlation between the RMSE$_{OBS-FOR}$ and the median and between RMSE$_{OBS-FOR}$ and $\sigma_{1h}$ on a rich statistical sample. These values are associated to Fig. \ref{RMSE_vs_mean_sigma}.}
\centering
\begin{tabular}{l c c}\toprule
Figure of merit & PP & PC\\ \hline
median: Pearson &  0.717 & 0.692 \\
std: Pearson &  0.887 & 0.866 \\ 
median: RMSE & 0.080 & 0.084\\
std: RMSE & 0.053 & 0.058\\ \hline
\end{tabular}
\label{table_correlation}
\end{table}

\section{Conclusions}
\label{conc}
This study has successfully quantified the reference performance thresholds for two baseline forecasting models: Precasting (PC) for predicting averages { values within intervals of time}, and Prevision by Persistence (PP) for { predicting continuous temporal evolution within interval of time}. { Such an analysis has been carried out above the site of VLT (at Cerro Paranal, Chile) and the site of LBT (At Mt. Graham, Arizona, US) where our team is leading two operative forecast systems (ALTA at LBT) and (FATE at VLT).} The analysis { has been focused on} short-term forecast timescales ($FTS = 1\text{h}$ and $2\text{h}$) { as these are most relevant time scales for science operations}. These models offer meaningful baseline thresholds; { all OT forecasting models that do not achieve a lower error than these baselines do not offer an added predictive value.} Because these thresholds are highly site-dependent, { it is highly recommended to include the estimates proposed in this study to appreciate the added value provided by a specific forecast method.}

These reference thresholds were derived for the primary astroclimatic parameters—seeing, isoplanatic angle, wavefront coherence time, and ground layer fraction (GLF)—utilising an extensive eight-year dataset from Cerro Paranal (VLT) and Mt. Graham (LBT). The performance profiles vary continuously depending on the specific parameter, geographical site, selected FTS horizon, and the specific FTS framework applied (Case A or Case B).

By expanding the analysis across seven distinct locations, this study also investigated the capacity to extrapolate baseline performance using local atmospheric statistics. The results demonstrate that while absolute site metrics like the median or mean seeing are poor predictors of baseline error, the standard deviation of the seeing { on the same time scale on which the forecast is performed (for example $\sigma_{1\text{h}}$ for a FTS=1h)} correlates with the $RMSE_{\text{FOR-OBS}}$. This implies that OT forecasting { is easier} for free-space optical communications, for which sites are typically characterised by less stable atmospheric conditions. In other words, in the FSOC field, there is a greater opportunity for added values from complex forecasting models. 

For a full discussion regarding the impact of resampling, dataset cleaning, and parallel trends for core meteorological parameters (WS, T, RH, PWV), the reader is referred to the full analysis in { de Sepibus et al.} \cite{desepibus2026}.

\begin{acknowledgements} 
We acknowledge the Large Binocular Telescope Observatory (LBTO), the European Southern Observatory (ESO), the Telescopio Nazionale Galileo (TNG), the Matera Space Center and the Miratlas company for access to their measurements.
\end{acknowledgements} 

\bibliographystyle{spiebib} 
\bibliography{references} 

\end{document}